\documentclass[aps,twocolumn,nobalancelastpage,tightenlines,floatfix]{revtex4}
\usepackage{amssymb}
\usepackage{graphicx}

\begin{document}

\title{Cooling to the Ground State of Axial Motion \\
for One Atom Strongly Coupled to an Optical Cavity}
\author{A. D. Boozer, A. Boca, R. Miller, T. E. Northup, and H.~J. Kimble}
\affiliation{Norman Bridge Laboratory of Physics 12-33\\
California Institute of Technology, Pasadena, CA 91125}

\begin{abstract}
Localization to the ground state of axial motion is demonstrated for
a single, trapped atom strongly coupled to the field of a high
finesse optical resonator. The axial atomic motion is cooled by way
of coherent Raman transitions on the red vibrational sideband. An
efficient state detection scheme enabled by strong coupling in
cavity QED is used to record the Raman spectrum, from which the
state of atomic motion is inferred. We find that the lowest
vibrational level of the axial potential with zero-point energy
$\hbar \omega_{\mathrm{a}}/2k_{B}=13~\mu K$ is occupied with
probability $ P_{0}\simeq 0.95$.
\end{abstract}

\date{\today}
\maketitle

Single atoms strongly coupled to the fields of high quality optical
resonators are of fundamental importance in Quantum Optics and, more
generally, can be used for many tasks in quantum information
science, including the implementation of scalable quantum
computation \cite{pellizzari95,duan04} and the realization of
distributed quantum networks \cite{cirac97,briegel00}. In recent
years, significant experimental progress to develop tools suitable
for these tasks has been made by employing optical forces to
localize individual atoms within optical cavities in a regime of
strong coupling
\cite{ye99,mckeever03a,maunz04,sauer04,boca04,nussmann05a,miller05},
as well as by combining trapped ions with optical cavities
\cite{ions}. Scientific advances thereby enabled include the
observation of the vacuum-Rabi spectrum for an individual atom
\cite{boca04} and vacuum-stimulated cooling \cite{nussmann05a}.

Although great strides are thus being made with atoms localized and
strongly coupled to the fields of optical cavities \cite{localize},
it has not previously been possible to access the quantum regime for
the atomic center-of-mass motion in cavity QED. Qualitatively new
phenomena have been predicted in this regime for which a quantized
treatment is required for both the internal (i.e., the atomic dipole
and cavity field) and external (i.e., atomic motion) degrees of
freedom, as was first recognized in the seminal work in Refs.
\cite{englert91,haroche91,holland91} and in the years since \cite
{storey92,herkommer92,ren95,scully96,vernooy97,doherty98,parkins99}.
Examples include the exploitation of strong coupling for the
reliable transfer of quantized states of atomic motion to quantum
states of light, and conversely \cite{parkins99}, as well as for
measurements that surpass the standard quantum limit for sensing
atomic position \cite{storey92}.

Our effort towards quantum control of atomic motion in cavity QED
unabashedly follows the remarkable set of achievements for trapped
ions \cite{leibfried03} and atoms in optical lattices
\cite{hamann98}, for which such control has led to the creation of
manifestly quantum (i.e., nonclassical) states of motion and to the
manipulation of quantum information. A first, enabling step in many
of these investigations has been the capability to cool to the
ground state of motion for single trapped atoms or ions.

\begin{figure}[tb]
\includegraphics[width=6.5cm]{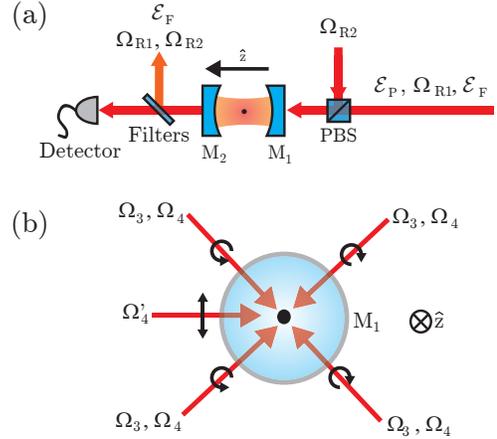}
\caption{Schematic of the experiment. The cavity is represented (a)
from the side and (b) along its axis. Shown are the various beams
used in the experiment: linearly polarized probe $\mathcal{E}_p$,
FORT $\mathcal{E}_F$, pumping $\Omega^{\prime}_4$, and Raman beams
$\Omega_{R1,R2}$, as well as the circularly polarized lattice beams
$\Omega_{3,4}$.} \label{fig:experiment-schematic}
\end{figure}

In this Letter we report localization to the ground state of motion
for one atom trapped in an optical cavity in a regime of strong
coupling \cite{miller05}. Resolved sideband cooling to the ground
state is accomplished with a coherent pair of intracavity Raman
fields. To deduce the resulting state of atomic motion, we introduce
a new scheme for recording Raman spectra by way of the strong
interaction of the atom with a resonant cavity probe. Our scheme is
the cavity QED equivalent of state detection in free space by
quantum-jump spectroscopy \cite{leibfried03}, and achieves a
confidence level for state discrimination $>98\%$ in
$100~\mathrm{\mu s}$. From the Raman spectra, we infer that the
lowest vibrational level $n=0$ of the axial potential is occupied
with probability $P_{0}\simeq 0.95$ for one trapped atom.

A schematic of the experimental setup is given in Fig.~\ref
{fig:experiment-schematic}. At the heart of the system is the
Fabry-Perot cavity formed by mirrors $(M_1,M_2)$. The cavity length
is stabilized to $ l_{0}=42.2~\mathrm{\mu m}$ using an independent
locking laser, such that a $TEM_{00}$ mode is resonant with the
$6S_{1/2},F=4\rightarrow 6P_{3/2},F^{\prime}=5^{\prime}$ transition
of the $D2$ line in Cs. The resulting atom-cavity coupling gives a
maximum single-photon Rabi frequency of $2g_0/2\pi =68~\mathrm{MHz}$
for $(F=4,m_F=\pm 4)\rightarrow
(F^{\prime}=5^{\prime},m_F^{\prime}=\pm 5)$. The decay rates are
$\gamma /2\pi =2.6~\mathrm{MHz}$ for the $6P_{3/2}$ excited states,
and $\kappa /2\pi =4.1~\mathrm{MHz}$ for the cavity field. Because
$g_0 \gg (\gamma ,\kappa)$, our system is in the strong coupling
regime of cavity QED \cite{miller05}, with critical photon and atom
numbers $n_0\equiv \gamma ^2/(2g_0^2)\approx 0.0029$ and $N_0\equiv
2\kappa \gamma /g_0^2\approx 0.018$.

Atoms are trapped by an intracavity far-off-resonant trap (FORT) at
$\lambda_{F}=935.6~\mathrm{nm}$, which is driven by a linearly
polarized input field $\mathcal{E}_{F}$ and is resonant with a
$TEM_{00}$ mode of the cavity with linewidth $\kappa_{F}/2\pi
=0.8~\mathrm{GHz}$. At $\lambda_{F}$, states in the ground $F=3,4$
and excited state $F^{\prime }=5^{\prime }$ manifolds experience
nearly-equal trapping potentials. For states in the $F=3,4$
manifolds, this potential is independent of $m_{F}$ and has a peak
value of $ U_{F}/h=-41~\mathrm{MHz}$, while for states in the
$F^{\prime }=5^{\prime }$ manifold it has a weak dependence on
$m_{F}^{\prime }$ \cite{mckeever03a,boca04}. The standing-wave
structure of the FORT forms independent wells where atoms may be
trapped. Near the bottom of a FORT well the potential is
approximately harmonic, with vibrational frequencies
$\omega_{a}/2\pi =530~\mathrm{kHz}$ for axial motion and
$\omega_{r}/2\pi =4.5~\mathrm{kHz}$ for radial motion.

To load atoms into the FORT, we release a cloud of cold atoms
located $\sim 3 $~mm above the cavity \cite{mckeever03a}. As the
atoms fall through the cavity, we apply $5~\mathrm{ms}$ of
$\Omega_{3}$, $\Omega _{4}$ light by way of two pairs of
counterpropagating $\sigma_{+}-\sigma_{-}$ polarized beams, as
illustrated in Fig.~\ref{fig:experiment-schematic}(b). These beams
are blue detuned by +10 MHz from the $F=3\rightarrow F^{\prime
}=3^{\prime }$ and $F=4\rightarrow F^{\prime }=4^{\prime }$
transitions, and cool the falling atoms via polarization gradient
cooling \cite{boiron96}. We adjust the powers \cite{powers} of these
beams so that the probability of loading at least one atom is $\sim
0.3$ per cloud drop.

Raman coupling between the $F=3$ and $F=4$ manifolds is generated by
driving a cavity mode at $\lambda_{R}=945.6~\mathrm{nm}$ with a pair
of beams $ \Omega_{R1,R2}$ that are phase-locked, lin $\perp $ lin
polarized, and have a relative detuning
$\omega_{R1}-\omega_{R2}=\Delta_{HF}^{\prime }+\delta_{R}$, where
$\Delta_{HF}^{\prime }/2\pi =9.19261~\mathrm{GHz}$ is the Cs
hyperfine splitting \cite{fortsqueezing} and $\delta_{R}$ is the
Raman detuning. This cavity mode has a linewidth $\kappa_{R}/2\pi
=6~\mathrm{GHz}$ , and the Raman beams are tuned such that
$\Omega_{R1(R2)}$ lies $\Delta_{HF}^{\prime }/2$ above (below)
cavity resonance. Since $\Omega_{R1,R2}$ drive a different mode of
the cavity than $\mathcal{E}_{F}$, atoms trapped in different FORT
wells see different Raman powers. If the relative spatial phase of
the FORT and the Raman pair at the bottom of a given well is $
\alpha $, then an atom at this potential minimum sees a Raman power
proportional to $\cos ^{2}\alpha $.

We typically set the optical power transmitted on resonance through
the cavity for $\Omega_{R1,R2}$ to $P_{R1}=P_{R2}=140~\mu $W, which
gives a Rabi frequency \cite{rabifrequency} $\Omega_{0}/2\pi
=200~\mathrm{kHz}$ for atoms with $\alpha =0$. The ac-Stark shift
due to these Raman beams adds a correction to the FORT potential of
$U_{R}/2\pi =0.84~ \mathrm{MHz}$. To avoid heating the atom by
switching $U_{R}$, we leave $ \Omega_{R1,R2}$ on all the time, but
usually keep them far-detuned ($\delta_{R}/2\pi =85~\mathrm{MHz}$)
from Raman resonance. To drive Raman transitions, we change $\omega
_{R2}$ so as to bring the pair into Raman resonance, whereas to
fine-tune $\delta_{R}$ we vary $\omega_{R1}$.

\begin{figure}[tb]
\centering
\includegraphics[width=7cm]{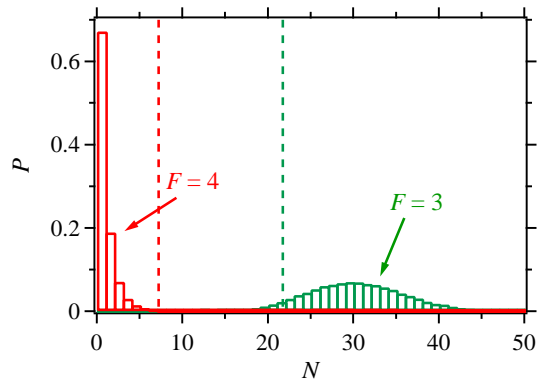}
\caption{Histogram of counts recorded during 100
$\mathrm{\protect\mu s}$ probing intervals for $N_e = 30$. The red
(green) curve is the probability $P $ of detecting a given number of
counts $N$ for an atom prepared in the $F=4$ ($F=3$) state. The
dashed vertical bars indicate detection thresholds at $N=0.25 N_e$
and $N=0.75 N_e$.} \label{fig:count-histogram}
\end{figure}

Because the intensity of the Raman pair is spatially varying, the
Raman coupling can connect states with different vibrational quantum
numbers. The form of this motional coupling is simple for atoms near
the bottom of a FORT well where the axial and radial motions
decouple, allowing us to consider the effect of the Raman coupling
on the axial motion alone. In this harmonic limit, we can define a
set of Fock states $\{|n\rangle\}$ for the axial motion. For
transitions from $F=3,m=0$ to $F=4,m=0$ and to first order in $
\eta$, the Rabi frequency for an $n\rightarrow n$ transition is $
\Omega_{n\rightarrow n} = (1/2)(1 + \cos 2\alpha)\,\Omega_0$, while
for an $ n\rightarrow n-1$ transition, it is $\Omega_{n\rightarrow
n-1} = \eta\sqrt{n}\sin 2\alpha\,\Omega_0$ (similar results hold for
other Zeeman sublevels \cite{boozer_prep}); here, $\eta =
(2\pi/\lambda_R) z_0 = 0.056$ is the Lamb-Dicke parameter for axial
motion, and $z_0 = \sqrt{\hbar/2m\omega_a}=8.5~\mathrm{nm}$ is the
ground state wavepacket size. Note that the $n\rightarrow n-1$
transition is strongest for atoms with $\alpha = \pi/4$.

The spatial dependence of the Raman coupling, together with the fact
that the the axial motion of the atom is in the Lamb-Dicke limit,
allows us to implement Raman sideband cooling \cite{leibfried03}. We
tune the Raman pair to the red axial sideband ($\delta_R = -525$ kHz
$\simeq -\omega_a$) \cite{signconvention} and apply the $\Omega_4$
lattice beams. An atom that starts in $F=3$ is coherently
transferred by $\Omega_{R1,R2}$ to $F=4$, where it is incoherently
repumped to $F=3$ by $\Omega_4$. The coherent transfer lowers the
axial vibrational quantum number $n$ by one, while the incoherent
repumping usually leaves $n$ unchanged since $n$-changing
transitions are Lamb-Dicke suppressed. Thus, the beams continually
lower $n$, cooling the atom to the axial ground state. Also, the
$\Omega_4$ light provides Sisyphus cooling \cite{boiron96} in the
radial direction.

Strong atom-cavity coupling enables a versatile detection scheme for
determining if an atom is present in the cavity, and if so, if it is
in the $ F=3$ or $F=4$ manifold. For the current settings, in
$100~\mathrm{\mu s}$ we measure the atomic hyperfine state with a
confidence level of $>98\%$. The scheme involves driving the cavity
with a $100~\mathrm{\mu s}$ pulse of resonant $F=4\rightarrow
F^{\prime }=5^{\prime }$ linearly polarized probe light
$\mathcal{E}_{P}$. As shown in Fig. \ref{fig:count-histogram}, if an
$ F=3$ atom is present (or if the cavity is empty) then the light is
transmitted, while if an $F=4$ atom is present then the light is
blocked because of the strong atom-cavity coupling \cite{boca04}.
Suppose we set the $ \mathcal{E}_{P}$ intensity such that on average
$N_{e}$ photons are detected \cite{detection-efficiency} per probing
interval when no atom is in the cavity. Then if the number $N$ of
detected photons is such that $N<0.25N_{e}$, we assume an $F=4$ atom
is present, if $N>0.75N_{e}$ we assume an $F=4$ atom is not present,
otherwise the measurement is inconclusive (which happens $<2\%$ of
the time) and we ignore the result. Whenever we detect the atomic
state we perform two such measurements: the first with just
$\mathcal{E}_{P}$ to find out if an $F=4$ atom is present; the
second with the $\mathcal{E}_{P}$ probe together with $\Omega_{3}$
as a repumper \cite{powers}, to see if an atom is present at all,
regardless of its internal state.

We measure the Raman transfer probability $P_4$ for a given
$\delta_R$ by preparing an atom in $F=3$, applying a Raman pulse,
and then detecting the atomic state using the above scheme (with
$N_e\sim 22$). We call one such measurement cycle a trial. For each
trial, we first Raman-sideband cool the atom for an interval $\Delta
t_c$. Next, we pump it into $F=3$ by alternating $1~\mathrm{\mu s}$
pulses of $\Omega_4$ lattice light with $1~\mathrm{\mu s}$ pulses of
$ \Omega^{\prime}_4$ linearly polarized resonant $F=4 \rightarrow
F^{\prime}=4^{\prime}$ light from the side of the cavity (10 pulses
of each). After the atom is pumped to $F=3$, we apply a $\Delta t_R
= 500~ \mathrm{\mu s}$ Raman pulse, which sometimes transfers it to
$F=4$. Finally, we measure the atomic state and check if the atom is
still present. For each atom, we fix the absolute value of the Raman
detuning $|\delta_R|$, and alternate trials at $+|\delta_R|$ with
trials at $-|\delta_R|$ (299 trials each). By combining data from
atoms with different values of $|\delta_R|$, we map out a Raman
spectrum. Note that because the initial Zeeman state of the atom is
random, all allowed $F=3 \rightarrow F=4$ Zeeman transitions
contribute to these spectra.

\begin{figure}[tb]
\centering
\includegraphics[width=6.5cm]{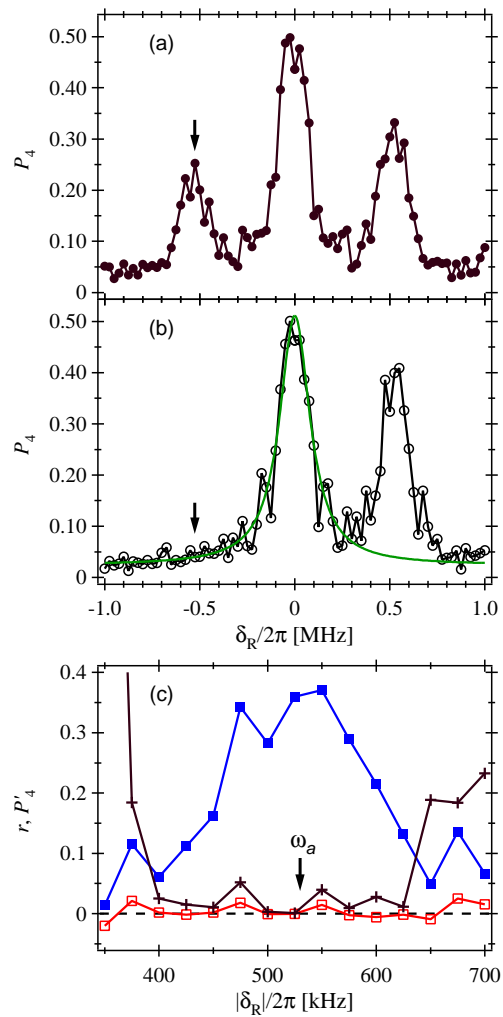}
\caption{Population $P_4$ in the $F=4$ state versus Raman detuning
$\protect \delta_R/2\protect\pi$. The (a) data are taken with
$\Delta t_c=250~\mathrm{ \ \protect\mu s}$ of cooling, and with an
$\Omega_4$ total 4-beam intensity $ I_4 = 5I_4^{\mathrm{sat}}$; the
(b) data with $\Delta t_c =5~\mathrm{ms}$ of cooling, and
$I_4=0.5I_4^{\mathrm{sat}}$ (on average, about 33 atoms per data
point). The arrow marks the detuning used for sideband cooling. (c)
Zoom-in on the two sideband regions for the (b) data, with detuning
axis folded around $\protect\delta_R = 0$. The red ($\Box $) and
blue ($ \blacksquare$) sidebands, as well as their ratio $r$ ($+$),
are shown after subtracting a Lorentzian fit to the carrier (green
curve in (b)).} \label{fig:raman-scans}
\end{figure}

Two example Raman spectra are plotted in Fig. \ref{fig:raman-scans}.
For the (a) curve, we cool for $\Delta t_c = 250~\mathrm{\mu s}$,
for the (b) curve for $\Delta t_c = 5~\mathrm{ms}$. These scans are
performed after nulling the magnetic field to within $\sim
40~\mathrm{mG}$; the widths of the peaks are set by the splitting of
different Zeeman levels due to the residual magnetic field. For the
curve in panel (a), we see peaks at the carrier ($ \delta_R = 0$),
as well as at the blue/red sidebands ($\delta_R/2\pi \simeq \pm
530~\mathrm{kHz} = \pm \omega_a$). Already we note a sideband
asymmetry, indicating that a significant fraction of the population
is in the $n=0$ vibrational state. For the (b) data, the red
sideband at $\delta_R/2\pi \simeq -530~\mathrm{kHz}$ is suppressed
to such a degree that it cannot be distinguished from the background
and contribution from off-resonant excitation of the carrier.

The ratio $r$ of transfer probabilities for the red and blue
sideband gives information about the temperature of the atom. For a
two-state atom in a thermal state, this ratio $r_0$ at $|\delta_R| =
\omega_a$ is related to the mean vibrational quantum number
$\bar{n}$ by $r_0 = \bar{n}/(\bar{n}+1)$ \cite{leibfried03}. In
Fig.~\ref{fig:raman-scans}c, we plot $r$ as a function of
$|\delta_R|$ for the $\Delta t_c=5~\mathrm{ms}$ data. As shown in
Fig.~\ref{fig:raman-scans}b, we fit a Lorentzian curve to the
carrier, then subtract its contribution from both the red and the
blue sideband data, with the result shown in panel (c). We find $r_0
\simeq \bar{n}= 0.01 \pm 0.05$, and the ground state population $P_0
= 1/(\bar{n}+1) =0.99 \pm 0.05$, where the error bars reflect
fluctuations in the data around $ |\delta_R|=\omega_a$. If instead
we subtract the constant background of $ P_4^B=0.024$ but not the
carrier's Lorentzian tail, we find $r_0 \simeq \bar{ n} = 0.05 \pm
0.04$, and $P_0= 0.95 \pm 0.04$. Finally, if we use the raw data
from Fig. \ref{fig:raman-scans}b with no subtractions, we obtain
$r_0 = 0.10 \pm 0.03$, $\bar{n} = 0.12 \pm 0.04$ and $P_0 = 0.89 \pm
0.03$. Note, however, that because the atom is not a two-state
system and the motional state is not known to be thermal, these
estimates are approximate.

\begin{figure}[tb]
\includegraphics[width=6.5cm]{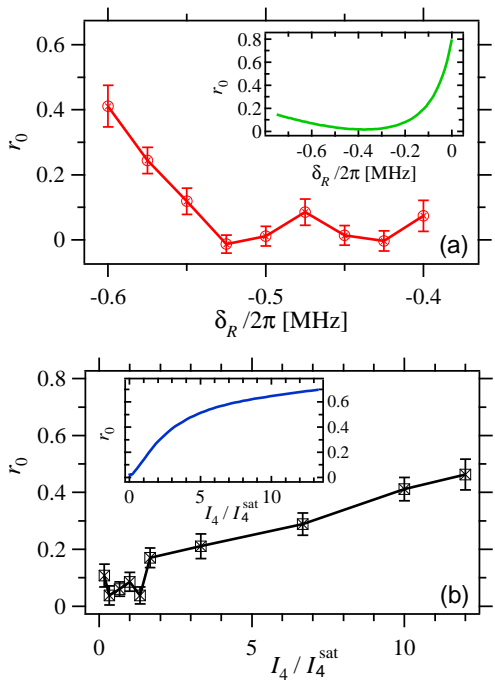}
\caption{Varying cooling parameters. The sideband ratio $r_{0}$ is
shown as a function of (a) the Raman detuning $\protect\delta_R$
employed for cooling and (b) the $4\rightarrow 3$ repumping
intensity $I_4$ \protect\cite{settings}. Insets show the results
from a simple calculation for a 2-state atom trapped in a FORT well
with $\protect\alpha=\protect\pi/4$.} \label{fig:vary_stuff}
\end{figure}

The axial cooling rate and asymptotic value of $\bar{n}$ depend on
$\delta_{R}$, on the $\Omega_{R1,R2}$ Rabi frequencies, and on the
power and detuning of $\Omega_{4}$. We have performed detailed
computer simulations to help us choose optimal values for these
parameters \cite{boozer_prep}. A common feature of both our
theoretical and experimental investigations is the remarkable
robustness of $\bar{n}$ under variations of the cooling parameters.
As an example \cite{settings}, in Fig.~\ref{fig:vary_stuff} we plot
the measured sideband ratio $r_{0}$ at $\delta_{R}/2
\pi=-500~\mathrm{ kHz} \simeq -\omega_{a}$ as a function of (a) the
detuning $\delta_{R}$ used for sideband cooling, and (b) the
recycling intensity $I_{4}$. The sideband asymmetry is maintained
over a range of at least $200$~kHz in detuning, and of two orders of
magnitude in the intensity $I_{4}$ of the $\Omega_{4}$ beams. The
insets give results from a simple 2-state calculation of $r_{0}$,
displaying similar insensitivity to the exact values of $\delta
_{R}$ and $I_{4}$ used for Raman sideband cooling
\cite{boozer_prep}.

As for the radial temperature, we do not yet have an accurate thermometer.
We can, however, adiabatically ramp down the FORT depth to zero, so that
only the $U_{R}$ trapping potential remains, and measure the probability
that the atom survives the ramping process \cite{alt03}. Assuming
three-dimensional thermal equilibrium (which might not hold for our system),
these data limit the temperature to $\sim 100~\mu \mathrm{K}$. However, the
Sisyphus cooling we use radially has been previously shown to reach
temperatures of $\sim 1~\mu \mathrm{K}$ \cite{boiron96}.

Another way to investigate the efficacy of the cooling is to examine its
impact on the lifetime of the atom in the trap. In the presence of state-
and atom-detection probing trials, the atomic lifetime is $0.4~\mathrm{s}$
with no cooling, $1.8~\mathrm{s}$ with $250~\mathrm{\mu s}$ of cooling
(settings as in Fig. \ref{fig:raman-scans}a), and $4.6~\mathrm{s}$ with $5~
\mathrm{ms}$ of cooling per trial (as in Fig. \ref{fig:raman-scans}b).

The Raman coupling provides a versatile tool for cavity QED with
trapped atoms beyond cooling the atomic motion that we have focused
on here. For example, we have used Raman spectroscopy to null the
magnetic field at the trapping sites to $\sim 10$ mG (including any
FORT pseudo-field), to prepare coherent superpositions of two ground
states, and to measure the population distribution among Zeeman
sublevels for various protocols \cite{boozer_thesis}.

In conclusion, we have demonstrated cooling to the ground state of
axial motion for single Cesium atoms strongly coupled to the field
of a small optical resonator. Together with existing capabilities
for strong coupling of the internal degrees of freedom, control over
the external center-of-mass motion in cavity QED enables a new set
of phenomena to be explored at the light-matter interface. For
example, arbitrary states of atomic motion can be prepared from the
ground state by coherent Raman transitions \cite{leibfried03}, then
mapped to the electromagnetic field by way of the strong atom-field
coupling \cite{parkins99}. Investigations of sensing atomic motion
at the standard quantum limit and feedback control now become
feasible \cite{storey92,doherty99}.

This research is supported by the National Science Foundation and by the
Disruptive Technology Office of the Department of National Intelligence. RM
acknowledges support from the Army Research Office (ARO).


\begin{thebibliography}{99}
\bibitem{powers} For trap loading, the total power in the four lattice beams
is $50I_{4}^{\mathrm{sat}}$ for $\Omega_{3}$ and $\Omega_{4}$, where
$ I_{4}^{\mathrm{sat}}\sim 3.8~\mathrm{mW}/\mathrm{cm}^{2}$. For
atom detection, the $\Omega_{3}$ intensity is
$5I_{4}^{\mathrm{sat}}$.
\bibitem{fortsqueezing} $\Delta_{HF}^{\prime }$ for a trapped atom is
slightly reduced compared to its free space value
$\Delta_{HF}=9.19263$~GHz, because the FORT potential for the $F=3$
manifold is slightly weaker than for $F=4$. For our FORT, the
correction is $\Delta_{HF}^{\prime }-\Delta_{HF}=-21~\mathrm{kHz}$
\cite{boozer_thesis}.
\bibitem{rabifrequency} Here $\Omega_0$ is the Rabi frequency for the
transition $F=3,m=0,n=0 \rightarrow F=4,m=0,n=0$, where the quantization
axis is along the cavity axis.
\bibitem{signconvention} Note that for $\delta_{R}=-\omega_{a}$, an $
F=3\rightarrow F=4$ ($F=4\rightarrow F=3$) Raman transition lowers
(raises) $ n$ by one.
\bibitem{detection-efficiency} The probability for a photon starting out
within the cavity to be recorded by our detectors is $e\sim 0.06$.
\bibitem{settings} Unless otherwise noted, the settings for this measurement
are: $I_{4}=0.3I_{4}^{\mathrm{sat}}$, $\Delta t_{c}=5$ ms,
$\delta_{R}=-500$ kHz, $\Delta t_{R}=50~\mu \mathrm{s}$;
$P_{4}^{B}=0.065$ was subtracted before computing $r_{0}$.
\end{thebibliography}
\end{document}